# Blue-enhanced Supercontinuum Generation in a Fluorine-doped Graded-index Multimode Fiber

Z. Sanjabieznaveh[(1,a)], M. A. Eftekhar[(1)], J. E. Antonio Lopez[(1)], H. Lopez Aviles[(1)], M. Kolesik[(2)], F. W. Wise[(3)], D. N. Christodoulides[(1)], and R. Amezcua Correa[(1)]

[1]CREOL, The College of Optics and Photonics, University of Central Florida, Orlando, Florida 32816, USA
[2]College of Optical Sciences, University of Arizona, Tucson, AZ 85721, USA
[3]School of Applied and Engineering Physics, Cornell University, Ithaca, New York 14853, USA
[a]zahoora@knights.cuf.edu

**Abstract**

Blue-enhanced, broadband supercontinuum generation in a 50 μm core fluorine-doped graded-index multimode fiber is demonstrated by pumping with a 1064 nm picosecond source. Multi-octave supercontinuum spectrum extending from ~400 nm to 2,400 nm is achieved in a relatively short fiber. The measured spectrum and multimode nonlinear properties are compared to those of a germanium-doped graded-index multimode fiber. It is shown that engineering the glass composition and thus its chromatic dispersion and absorption characteristics, provides a novel and effective approach to further controlling the nonlinear dynamics in highly multimoded parabolic-index fibers.

**Introduction**

Supercontinuum (SC) generation in optical fibers has been a subject of intense interest in recent years. Owing to its broad spectral content and high spatial brightness, SC light has potential applications in spectroscopy, fluorescence microscopy, biomedicine and optical telecommunications[1]. Thus far, photonic crystal fibers (PCFs) have been particularly successful in generating efficient SC as they have flexible dispersion characteristics and high nonlinearity resulting from small mode area[1–9]. To enhance the visible range of the SC spectrum, tapered PCFs[6,10], and PCFs with tailored dispersion characteristics[7] have been employed. Nevertheless, generation of efficient SC in the blue wavelength range in PCFs has been quite challenging because it requires an engineered zero dispersion wavelengths and a suitable phase-matched pump wavelength[7]. Besides, the small core size of the PCF, ultimately limits the maximum spectral power density that can be delivered during SC generation.

Recently, multimode fibers (MMFs) have attracted considerable attention for optical communications systems[10–13], and high power fiber lasers[14,15]. Owing to the multitude of possible nonlinear interactions among their guided modes, MMFs can open new directions for the development of fiber-based light sources. The broad SC generation in graded-index (GI) MMFs is explained based on the periodic spatial beam refocusing which leads to geometric parametric instability (GPI)[16]. As a high-intensity pulse propagates in a GI-MMF, GPI efficiently induces a series of intense frequency components in the visible and near-infrared (NIR) spectral regions which can eventually evolve into an ultra-broadband SC[17–22].

Thus far, all reports of SC in GI-MMFs has been exclusively performed in germanium-doped GI-MMFs. Particularly, Lopez-Galmiche et al. demonstrated visible to NIR SC generation by pumping a germanium-doped GI-MMF in the normal dispersion regime[20]. However, in this experiment, long fiber length (>25 m) was required in order to extend the SC into the visible. While modifying the glass composition of PCFs has been reported to enhance the visible region of the SC[23], almost no effort has been devoted so far to controlling the material composition of the GI-MMFs in order to blue-shift the SC spectrum.

Here, we present a blue-enhanced SC generation in a 10 m long, fluorine-doped GI-MMF, pumped at 1064 nm. The fluorine-doped silica glass is exploited to further extend the short-wavelength edge of the SC towards the blue region[24,25]. The spectral evolution of the continuum is recorded as a function of pump power. In addition, the GPI-induced spectrum in this fiber is compared to that of a ~28.5 m long, germanium–doped GI-MMF reported recently[20]. Our results indicate that dispersion engineering of GI-MMFs through suitable choice of glass composition, can be a promising approach for realizing flat optical sources covering the entire visible spectrum.

**Results**

The single mode output of an amplified Q-switched microchip laser at 1064 nm was coupled into the 50 μm core, fluorine-doped GI-MMF. The maximum refractive index depression of the fiber was ~-16×10$^{-3}$ with respect to the pure silica. The energy, duration and the repetition rate of the input pulses were 95 μJ, 400 ps and 500 Hz, respectively. A half-wave plate and a polarizing beam splitter were used to control the power launched into the fiber. The free-space coupling efficiency was higher than 85% using a 50 mm focal length lens. Light at the output of the MMF under test was analyzed using two optical spectrum analyzers (OSAs), covering the spectral range from 350 nm to 1750 nm (ANDO AQ 6315E) and 1200 nm to 2400 nm (Yokogawa AQ6375).

Figure 1 compares a typical image of the dispersed output spectrum of a 10 m long fluorine-doped GI-MMF (Fig. 1(a)) and a ~28.5 m long germanium-doped GI-MMF[20] (Fig. 1(b)) pumped at a peak power of ~185 kW. Clearly, the visible SC in the fluorine-doped MMF shows a rather uniform spectrum from blue to red spectral regime. In addition, the far field image of the output beam, exhibits a bright white-light beam profile. In contrast, the visible spectrum of the germanium-doped GI-MMF, with an equal core size (50 um), similar refractive index profile (index suppression of ~16×10$^{-3}$ with respect to silica), and under the same pumping condition, displays several discrete peaks in the blue spectral regime and a yellowish output beam profile (Fig. 1(b)).

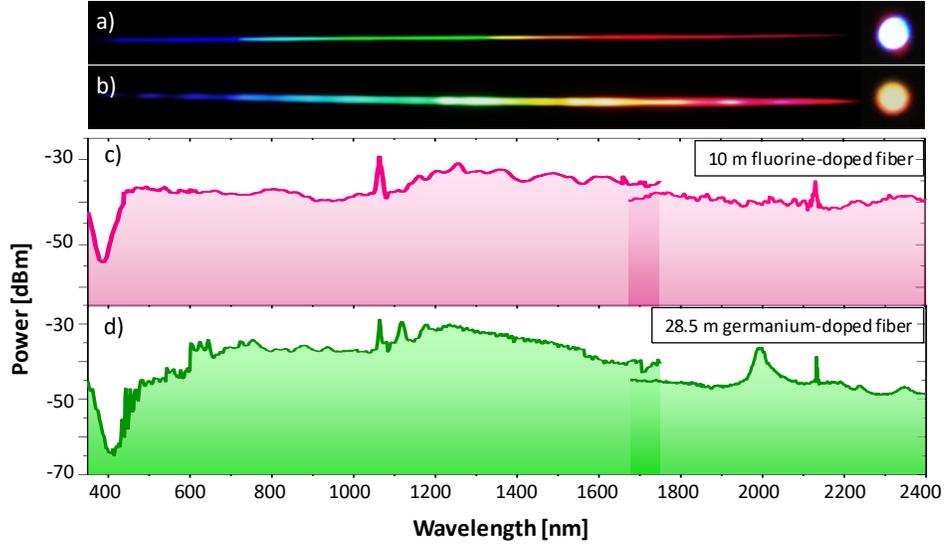

**Figure 1**. Comparison of SC generation in fluorine-doped and germanium-doped GI-MMFs. Dispersed output spectrum and far-field intensity profile of a (a) 10 m long fluorine-doped GI-MMF and (b) ~28.5 m long germanium-doped GI-MMF presented in[20]. SC spectrum in (c) fluorine-doped and (d) germanium-doped GI-MMFs pumped at 1064 nm at 185 kW peak power.

The generated SC spectrum in the fluorine-doped and germanium-doped GI-MMFs are shown in Fig. 1 (c) and (d), respectively. These spectra were recorded using two different OSAs at operating wavelength range of 350 nm-1750 nm and 1700 nm- 2400 nm, respectively. As it is shown, the fluorine-doped fiber displays a rather smooth spectrum in the visible range, with intensity variations of less than 5 dB from 450 nm to 1000 nm. On the other hand, the SC spectrum of the germanium-doped fiber exhibits multiple sharp peaks in the visible, with intensity variations of >10 dB over the same frequency range (Fig. 1(d)). This behavior can be attributed to the different chromatic dispersion of the fibers under test as discussed in the following.

Figure 2 compares the dispersion curves of pure silica (green), germanium doped (blue) and fluorine doped (red) fibers, respectively. Refractive index profile of the both fibers are shown as inset to fig.2.

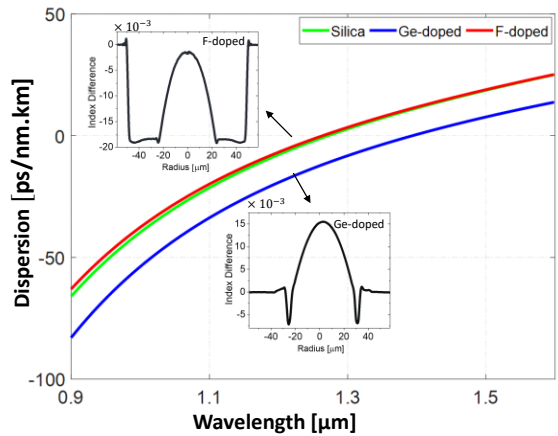

**Figure 2.** Dispersion curve of silica (green), germanium doped fiber (blue) and fluorine doped fiber (red). The refractive index profile of the fibers are shown in the inset.

As Fig.2 shows, the fluorine-doped fiber displays higher dispersion than the germanium-doped fiber. Therefore, pulse propagating in the fluorine-doped fiber broadens more rapidly than in the germanium-doped fiber. This results in the SC generation in a shorter length of fluorine-doped fiber in contrary to the longer length of the germanium-doped fiber, which is in perfect agreement with the results of Fig.1. In addition, fluorine-doping leads to the blue-shift of the silica absorption edge, improving the transparency in short wavelengths[24]. No sign of performance degradation was observed after long hours of operations (~ 24 hours) in either of the fibers.

To characterize the spatial intensity profile of the output beam of the fluorine-doped MMF, the near-field and visible far-field mode distributions were recorded using 10 nm bandpass filters. It can be seen from Fig. 3 that at the highest pump peak power levels (185 Kw), the spatial mode profiles are Gaussian-like across the entire SC spectrum ($M^2$< 1.8 across the whole spectral range). This can be attributed to pure Kerr nonlinear processes (intermodal cross-phase modulation and four-wave-mixing). This feature is of great importance for applications in high power delivery for which a stable, high-brightness beam profile output is desirable.

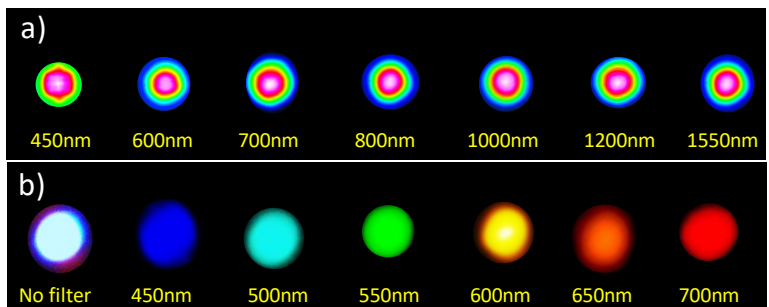

**Figure 3**. (a) Near-field and (b) visible far-field patterns of the beam profiles at the output of a 10 m fluorine-doped GI-MMF.

In order to clarify the nonlinear dynamics and the evolution of the continuum in our fluorine-doped GI-MMF, we have studied the spectral broadening as a function of the input pump power in a 10 m long fiber. As shown in Fig. 4, at low input pump peak power (~28 kW), several discrete bands around the pump wavelength can be clearly observed arising from Raman interactions and SPM.

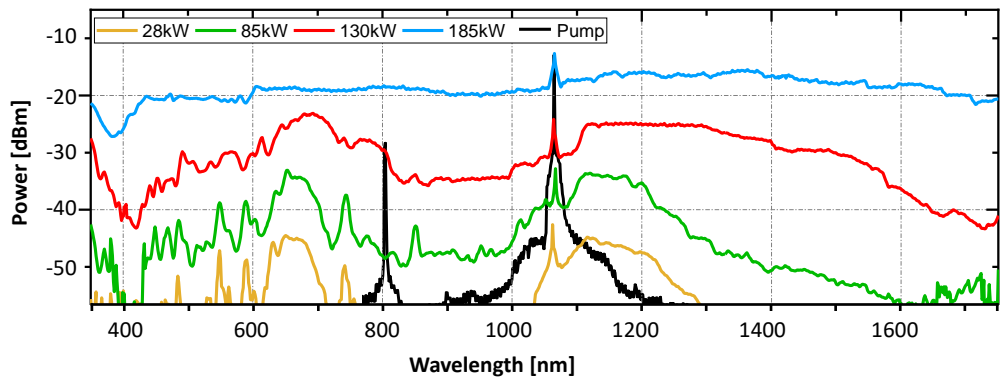

**Figure 4.** SC generation evolution as a function of pump power in a 10 m fluorine-doped GI-MMF. The input peak power is increased from 28 kW to 185 kW. The pump spectrum is shown in black.

At 85 kW, additional discrete frequency peaks appear in the spectrum. As the pump peak power was further increased to 130 kW, the spectrum starts to exhibit a continuous broadening on both sides of the pump wavelength. At even higher pump peak powers (~185 kW), extremely wide spectral broadening of the input pulse occurs, extending from ~400 nm to more than 1500 nm. In this figure, the initial pump spectrum is shown in black. It is to be noted that the peak at 806 nm is the residual of the initial pump power.

**Discussion**

To better understand the experimental observations in Fig.1, the numerical simulations were performed based on a general unidirectional pulse propagation equation (gUPPE) method[26]. In this approach, the spatiotemporal evolution of the total electric field in the presence of linear (dispersion, guiding) and nonlinear effects (SPM, FWM, third harmonic generation, shock and Raman effects) are computed at each step. In our simulations, the time window, the size of the spatial window and an adaptive integration step size were 6 ps, 50 µm × 50 µm and $1 - 2$ µm, respectively. In order to shorten the computational time, the propagation distance was set to 34 cm and the pulse width was taken to be 400 fs, with a beam waist of 27 µm.

Figure 5 compares the simulation results of the spectral evolution as a function of propagation distance in the fluorine-doped (Fig.5 (a)) and germanium-doped (Fig.5 (b)) GI-MMFs. This Figs. show that at the early stages of propagation, the spectral broadening in both fibers is mainly due to SPM. However, in fluorine-doped fiber, after about 5 cm of propagation, the down-conversion of the pump generates the first GPI induced NIR line at ~ 0.142 PHz (2100 nm). Subsequently, after ~ 10 cm, a series of sidebands emerge (extending from NIR to the visible) with the first ones located at 0.380 PHz (789 nm), 0.450 PHz (665 nm), 0.50 PHz (600 nm), 0.542 PHz (553 nm) in the visible and at 0.195 PHz (1500 nm) in the NIR. More spectral features eventually develop as a result of Raman and FWM, evolving in a broad SC light extending from below 0.15 PHz (2 um) to above 0.65 PHz (460 nm).

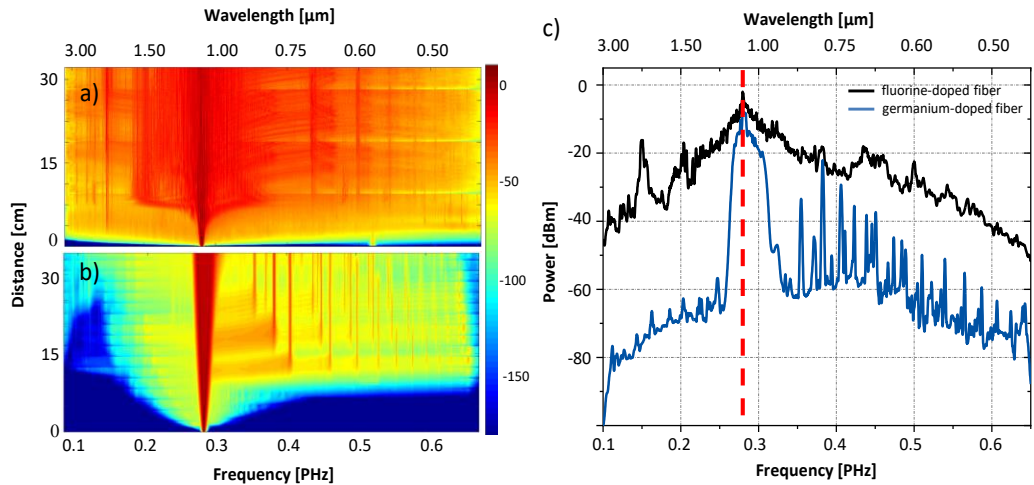

**Figure 5.** Spectral evolution upon propagation in a ~32 cm (a) fluorine-doped and (b) germanium-doped GI-MMFs with 50 µm core diameter. (c) Simulated spectra in the fluorine-doped (black curve) and germanium-doped (blue curves) MMFs. The red dotted line represents the position of the pump at 1064 nm.

In contrast, light propagating in the germanium-doped GI-MMF, induces a series of visible sidebands after ~10 cm (Fig. 5(b)), with the first line located at 0.414 PHz (724 nm). Figure 5(c) compares the frequency positions of the induced GPI sidebands in a 1.5 m of fluorine-doped (black) and germanium-doped (blue) GI-MMFs. As shown, while a broad spectrum is generated in fluorine-doped fiber, there are several sharp peaks in germanium-doped GI-MMF, in close agreement with the experimental observations. Consequently, the spatial broadening of a pulse propagating in a GI-MMF strongly depends on the material composition of the core. In this regard, fluorine-doped silica fiber is a potential glass composition to generating SC at the early stages of ps pulse propagation.

In summary, we experimentally demonstrated the generation of a blue-enhanced, broadband SC in a 10 m long fluorine-doped GI-MMF using 400 ps pump pulses at 1064 nm. The generated SC was compared to that of a 28.5 m long, germanium-doped GI-MMF with an equal core size and similar refractive index profile. While SC generation in the latter exhibits several sharp peaks in the range of short wavelengths, the SC in fluorine-doped fiber showed a quite flat spectrum within a broad wavelength range of 450-2400 nm. This feature is mainly attributed to different chromatic dispersion of fibers under test as well as the blue-shift of the abortion edge of the silica glass doped with fluorine. In addition, spectral broadening in the fluorine-doped fiber was shown to arise mainly from induced GPI effects and the combined action of SRS and FWM. Our experiments suggest that by controlling the dispersion characteristic of the core through appropriate choice of the fiber material composition, we can convert a NIR laser beam into a flat, ultra-broadband SC source covering the whole visible spectral range. Such sources could potentially find applications in biomedical imaging and visible-based microscopy and spectroscopy systems.

## Author contributions statements

Z. Sanjabi Eznaveh performed the experimental part of the paper.
M. A. Eftekhar, H. Lopez Aviles and M. Kolesik contributed to the numerical modeling and analytical equations.
J. E. Antonio Lopez drew the fluorine doped fiber using a fiber drawing tower.
Z. Sanjabi Eznaveh, and M. A. Eftekhar wrote the main manuscript text.
F. W. Wise, D. N. Christodoulides, and R. Amezcua Correa supervised the entire work.
 All authors reviewed the manuscript.

**ACKNOWLEDGMENT**

This work was supported by the Office of Naval Research (ONR) (MURI N00014-13-1-0649), HEL-JTO (W911NF-12-1-0450), Army Research Office (ARO) (W911NF-12-1-0450), National Science Foundation (NSF) (ECCS-1711230) and Air Force Office of Scientific Research (AFOSR) FA9550-15-10041. This Letter used the Extreme Science and Engineering Discovery Environment (XSEDE).